\documentclass[aps,amsmath,amssymb,twocolumn]{revtex4}
    \usepackage{graphicx}       \usepackage{bm}
    \usepackage{epsf}           \usepackage{epsfig}
\newcommand{\bra}[1]{\mbox{$\left\langle #1 \right|$}}
\newcommand{\ket}[1]{\mbox{$\left|#1\right\rangle$}}

\newcommand{\ignore}[1]{}
\begin{document}
\title{Topological Order Following a Quantum Quench}
\author{Dimitris I. Tsomokos$^{1,2,3}$, Alioscia Hamma$^{4}$, Wen Zhang$^{5}$,
Stephan Haas$^{5}$, Rosario Fazio$^{2,6}$}
\affiliation{$^{1}$School of Physics and Astronomy, University of
Hertfordshire, Hatfield AL10 9AB, UK \\
$^{2}$Scuola Normale Superiore, Piazza dei Cavalieri 7, 56126
Pisa, Italy \\
$^{3}$Department of Mathematics, Royal Holloway,
University of London, Egham, TW20 0EX, UK \\
$^{4}$Perimeter Institute for Theoretical Physics, 31 Caroline St.
N, N2L 2Y5, Waterloo ON, Canada \\
$^{5}$Department of Physics and Astronomy, University of Southern
California, Los Angeles, CA 90089, USA \\
$^{6}$Center for Quantum Technologies, National University of
Singapore, Republic of Singapore}
\date{\today}
\begin{abstract}
We determine the conditions under which topological order survives
a rapid quantum quench. Specifically, we consider the case where a
quantum spin system is prepared in the ground state of the Toric
Code Model and, after the quench, it evolves with a Hamiltonian
that does not support topological order. We provide analytical
results supported by numerical evidence for a variety of quench
Hamiltonians. The robustness of topological order under
non-equilibrium situations is tested by studying the topological
entropy and a novel dynamical measure, which makes use of the
similarity between partial density matrices obtained from
different topological sectors.
\end{abstract}
 \maketitle

Quantum states with topological order (TO) belong to non-symmetry
breaking phases of matter that cannot be characterized by local
order parameters~\cite{Wen_book}. The absence of any local
structure in these states is ultimately responsible for their
robustness against certain forms of local noise. For this reason
topologically ordered quantum systems with ground state degeneracy
and persistent energy gap, such as the Toric Code Model (TCM),
have been proposed as robust quantum memories~\cite{Kitaev}.
Moreover, systems with TO can support non-Abelian anyons as
quasiparticle excitations, in which case universal quantum
computation based on \emph{topological}, and thus more robust,
interactions is indeed possible \cite{TQC_review}.

Evaluating the persistence of topological order under adverse
conditions is a crucial step towards future experimental
realizations of topological quantum hardware. The robustness of
the TCM has recently been studied with respect to
noise~\cite{Nayak} and thermal equilibrium at a finite
temperature~\cite{Finite_T_1, Finite_T_2, Finite_T_3}. The quantum
phase transition from topological to magnetic order in the TCM has
also been analyzed~\cite{Hamma_detectionTQPT, Claudio, Vidal}. We
can say that a system possesses TO if it is gapped and it has a
degenerate ground state whose degeneracy depends on the system
topology. If topology is to be responsible for the degeneracy, it
is expected that different ground states should be
indistinguishable locally. We show here that this is certainly the
case for the TCM and use this property to characterize the
presence of TO. In the absence of a local order parameter,
establishing the presence of TO is a daunting task. An effective
order parameter for TO is the topological entanglement entropy
\cite{S_topo,Zanardi}, $S_{\rm topo}$: if $S_{\rm
topo}(\ket{\psi}) = 0$, the state $\ket{\psi}$ does not have TO;
while if $S_{\rm topo}(\ket{\psi})>0$, the state has TO. $S_{\rm
topo}$ has already given insights into the nature of the TCM and
its transitions to non-topological phases
\cite{Finite_T_2,Finite_T_3, Hamma_detectionTQPT, Claudio}. By
contrast, the dynamical properties of TO have remained largely
unexplored.

In this work we investigate the dynamical response of a state with
TO to a sudden quench, a topic that has attracted a lot of
attention recently both experimentally \cite{non_eq_exp} and
theoretically \cite{quenches}. If the dynamics of the system is
governed by a Hamiltonian that does not support TO then it is
crucial to determine the extent to which TO survives and to assess
the nature of destructive perturbations. In particular, we examine
the case of a \emph{fast quench} starting from the ground state of
the TCM \cite{footnote}. We determine the influence of such a
quench on the TO by studying the dynamical behaviour of the block
entanglement entropy \cite{foot_refs} and the topological
entanglement entropy, whose dynamics is calculated here for the
first time. In addition we introduce a new \emph{dynamical
measure} of TO, which is based on the similarity between states
obtained from different topological sectors.

We start by considering a system of spin-$\frac{1}{2}$ particles
situated on the edges of a square lattice with periodic boundary
conditions. The TCM Hamiltonian on such a lattice is $H_{\rm 0} =
-\sum_{s} A_{s} - \sum_{p} B_p$, where $A_{s} = \otimes_{l \in s}
X_{l}$ is a \emph{star} operator applied to spins on edges that
meet on a vertex and $B_{p} = \otimes_{l \in p} Z_{l}$ is a
\emph{plaquette} operator on edges that form a square ($X$ and $Z$
denote the usual Pauli operators). These operators stabilize the
ground state of the model, that is, $\ket{\Psi_0} = A_{s}
\ket{\Psi_0} = B_{p} \ket{\Psi_0}$. There is fourfold degeneracy
in the ground state corresponding to different topological sectors
\cite{Kitaev}. This is because  the incontractible strings along
the vertical and horizontal directions, denoted by $w_1$ and
$w_2$, commute with the Hamiltonian and map $\ket{\Psi_0} $ into
orthogonal sectors. The ground state manifold can thus be written
as $\mathcal L = \mbox{span}\{w_1^iw_2^j\ket{\Psi_0}; \;
i,j=0,1\}$ with $\ket{\Psi_0} =|G|^{-1/2} \sum_{g \in G} g
\ket{\bf{0}}$, where $g$ is a string of star operators and
$\ket{\bf{0}} \equiv \ket{00 \cdots 0}$ is a reference state
(vacuum) in which all the spins are pointing in the $+Z$ direction
(equivalently, one may choose to define the vacuum in the
$X$-basis and the strings $g$ in the $Z$-basis). The Abelian group
$G$ of all star operators on an $m \times n$ lattice with $N =
2mn$ spins is generated under tensor multiplication. Due to the
periodic boundary conditions, $|G| = 2^{mn-1}$.

The lattice is prepared in the ground state $\ket{\Psi_0}$ of the
TCM Hamiltonian $H_{\rm 0}$. At $t=0$ a different Hamiltonian $H$
is suddenly switched on. The evolution is then given by
$\ket{\Psi(t)} = \exp(-i H t) \ket{\Psi_0}$. If $\{\epsilon_l,
\ket{\phi_l}\}$ is the spectrum of $H$ then $\exp(-i H t) =
\sum_{l=1}^{2^N} \exp(-i \epsilon_l t) \ket{\phi_l} \langle \phi_l
|$ and so
\begin{eqnarray} \label{evolution_general}
\ket{\Psi(t)} = \frac{1}{\sqrt{|G|}} \sum_{l=1}^{2^N} \sum_{g \in
G} \langle \phi_l | g \ket{\bf{0}} \exp(-i \epsilon_l t)
\ket{\phi_l}.
\end{eqnarray}
This is the most general form of the quenched state.

Firstly we prove that there exist families of quench Hamiltonians
that cannot disturb TO. To this end we consider that the
eigenstates of $H$ can be written as general spin flips $x$ acting
on the vacuum, i.e., $\ket{\phi_l} = x \ket{\bf{0}}$. The spin
flips define a group ${\cal X}$ under tensor multiplication, $\{x
\in {\cal X}|x = X_i \otimes X_{j} \otimes \cdots \}$. To apply
the argument also to the group $\mathcal Z$ of all
$Z_l$-operators, we take the initial state to be in a general
superposition of the four topological sectors. Then the quenched
state of Eq.~(\ref{evolution_general}) becomes $\ket{\psi(t)}
=|G|^{-1/2} \sum_{g \in G,x_l\in \mathcal{X},i,j} \alpha_{ij}$
$\exp[-i \epsilon_l t] F_{ij}\ket{\phi_l}$ with $F_{ij} \equiv
\bra{\bf{0}}x_l w_1^i w^j_2 g \ket{\bf{0}}$. The only surviving
terms are those such that $x_l=w_1^i w^j_2 g\equiv g^{ij}$ and
therefore we obtain
\begin{eqnarray} \label{evolution_specific}
\ket{\psi(t)} = \frac{1}{\sqrt{|G|}} \sum_{g \in G, ij = 0,1}
\alpha_{ij} \exp[-i \epsilon(g^{ij}) t] g^{ij} \ket{\bf{0}}.
\end{eqnarray}

The entanglement between a contractible region $A$ of the lattice
and its complement, $B$, is given by the entanglement entropy $S_A
= - {\rm Tr}(\rho_A \ln \rho_A)$, where $\rho_A$ is the reduced
state corresponding to region $A$. Using the replica trick
\cite{entropy}, $S_A (t) = - \lim_{n\rightarrow 1} \partial_n {\rm
Tr} [\rho_{A}^{n}(t)]$, where $\rho_A(t) = {\rm Tr}_B
[\ket{\Psi(t)} \bra{\Psi(t)}]$ is calculated for the general
quenched state. Specifically for the state of Eq.
(\ref{evolution_specific}), we have $\rho(t) = |G|^{-1}
\sum_{g,\tilde{g} \in G, ijkl=0,1} \alpha_{ij}\alpha^*_{kl} {\cal
E}^{ijkl}(g,g\tilde{g}) g \ket{{\bf 0}} \bra{{\bf 0}} g\tilde{g}$,
where ${\cal E}^{ijkl}(g,\bar{g}) \equiv \exp[-i\epsilon(g^{ij})t]
\exp[i \epsilon(\bar{g}^{kl})t]$ and we have used the property
$\bar{g} = g\tilde{g}$ of the group $G$. In order to calculate the
state $\rho_A (t) = {\rm Tr}_B [\rho(t)]$ we distinguish the
region, $\ket{{\bf 0}} \equiv \ket{0_A} \otimes \ket{0_B}$, and we
also define the subgroups of $G$ that act trivially on the
subsystems, i.e., $G_A \equiv \{g \in G | g = g_A \otimes
\openone_B \}$ and $G_B \equiv \{g \in G | g = \openone_A \otimes
g_B \}$. Since $A$ is contractible, there are always group
elements $h_1,h_2\in G$ such that $h_i w_i \in G_B$ for $i=1,2$.
In other words, it is always possible to deform the loop $w_i$ so
as to move it out of the region $A$. Therefore, we have $\rho_A(t)
= |G|^{-1} \sum_{g \in G, \tilde{g} \in G_A, ij=0,1}|\alpha_{ij}
|^2 {\cal E}^{ijij}(g,g\tilde{g}) g_A \ket{0_A} \bra{0_A} g_A
\tilde{g}_A$. Notice that for the ground state of the TCM,
$\rho_A$ does not depend on the coefficients $\alpha_{ij}$,
meaning that two orthogonal ground states are locally identical.

The particular structure of the Hamiltonian $H$ implies that
$\epsilon(g^{ij}) = k ^{ij}\epsilon (g)$ where $k^{ij}$ is a
constant; hence $\rho^{n}_{A}(t) = |G|^{-n}$ $\sum_{g_l \in
G,\tilde{g}_l \in G_{A}} \Pi(t)$ $g_{1,A} \ket{0_A} \bra{0_A}
g_{1,A} \tilde{g}_{1,A}$ $g_{2,A} \ket{0_A} \bra{0_A} g_{2,A}
\tilde{g}_{2,A} \cdots g_{n,A} \ket{0_A} \bra{0_A} g_{n,A}
\tilde{g}_{n,A}$, where the $\Pi(t) \equiv {\cal
E}(g_1,g_1\tilde{g}_1)$ ${\cal E}(g_2,g_2\tilde{g}_2) \cdots {\cal
E}(g_n,g_n\tilde{g}_n)$ includes the time dependence of the state.
Each expectation value of the form $\bra{0_A} g_{l,A}
\tilde{g}_{l,A} g_{l+1,A} \ket{0_A}$ is set equal to $1$ and this
imposes the condition $g_{l+1,A} = g_{l,A}\tilde{g}_{l,A}$ and so
$g_{l} \tilde{g}_{l} g_{l+1} \in G_B$. In this way we get
$\rho^{n}_{A}(t) =$ $|G_B|^{n}|G|^{-n}$ $\sum_{g_{1,n} \in
G,\tilde{g}_n \in G_{A}}$ $\Pi(t)$ $g_{1,A} \ket{0_A} \bra{0_A}
g_{n,A} \tilde{g}_{n,A}$, where now $\Pi (t) = {\cal E}
(g_1,g_n\tilde{g}_n)$. Using the replica trick we find $S_A(t) =
\ln (|G| / |G_A| |G_B|)$. Therefore the entanglement entropy of
the quenched state $\ket{\psi(t)}$ of Eq.
(\ref{evolution_specific}) remains constant in time. Moreover this
state obeys an area law for the entanglement, $S = L_A -1$, where
$L_A$ is the length of the boundary of $A$ and the universal
finite correction signals the presence of TO \cite{Zanardi}.

For the topological entropy we use the expression due to Levin and
Wen \cite{S_topo}, $S_{\rm topo} = -S_{ABCD} + S_{ABC} + S_{ACD} -
S_{AC}$ where $A,B,C,D$ are the sides of a $2\times 2$ square in
the lattice each containing two spins \cite{Hamma_detectionTQPT}.
Evidently, $S_{\rm topo}$ is time-independent for the quenched
state of Eq. (\ref{evolution_specific}) since the block
entanglement entropy is constant for this state. This completes
the proof of the first result, which can be stated as follows:
$S_{\rm topo}(t) = 1$ if the eigenstates of the quench Hamiltonian
can be written as spin flips on the vacuum (in either basis). For
this reason we call such quenches \emph{non-destructive}. Any
Hamiltonian $H(Z)$ or $H(X)$ that depends on only one Pauli
operator has such eigenstates. Physically relevant cases include
$H_{1} = - \sum_{i = 1}^{N} h_i S_i$ and $H_{2} = - \sum_{\langle
i,j \rangle} J_i S_iS_j$, where $S=X,Z$ and $i,j$ are nearest
neighbors. In the absence of disorder (i.e., for $h_i = h$ and
$J_i = J$) we have $|\langle \Psi_0 | \psi(t) \rangle|$ $=
|G|^{-1} \left| \sum_{g \in G} \cos[E(g) t] \right|$, where $E(g)$
are the eigenenergies in either case. Using the commutativity of
terms in $H_{1,2}$ it is easy to show that such quenches have a
period of recurrence $T=\pi /2h, \pi/2J$ independently of the
lattice size.  For this reason, we say that such a quench is
\emph{trivial}. Moreover, the entanglement remains constant at all
times even in the disordered case. Conversely, a quench will be
non-trivial if the periodicity is lost in the large system limit
and the recurrence time increases with the system size. In order
to detect the periodicity of the system we will study the overlap
$|\langle \Psi_0 | \Psi(t) \rangle |$ between the initial state
$\ket{\Psi_0}$ and the quenched state of Eq.
(\ref{evolution_general}).

In order to detect how TO is preserved by a non-trivial quench we
use $S_{\rm topo}$ and, in addition, the Uhlmann fidelity ${\cal
F}(t)$ between reduced systems that correspond to orthogonal
ground states. The ${\cal F}(t)$ is given by ${\cal F}(t) = {\rm
Tr} \sqrt{\sqrt{\sigma^{ij}_A(t)} \sigma^{kl}_A(t)
\sqrt{\rho^{ij}_A(t)}}$, where $\sigma^{ij}(t=0)$ is the density
matrix of $w_1^iw_2^j\ket{\Psi_0}$. As we have seen above, in the
topological phase the reduced systems are indistinguishable and
${\cal F}(0) = 1$, by construction. In numerical calculations with
finite size systems we need to choose the region $A$ such that its
bulk contains more degrees of freedom than the boundary. For this
reason we use this measure only for maximum available lattice
sizes.


\begin{figure} \centering
\resizebox{0.8\linewidth}{!} {\includegraphics{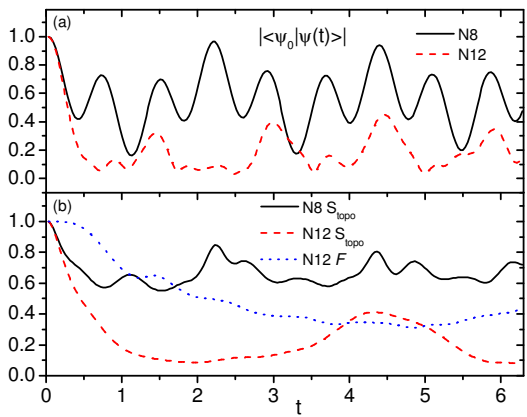}}
\caption{(Color online) Numerical results for the quench $H_3$
with couplings $J_1 / J_2 = 0.33$. (a) Absolute overlap $|\langle
\Psi_0 | \Psi(t) \rangle |$ on lattices with $N=8,12$ spins. (b)
$S_{\rm topo}(t)$ for the $N=8,12$ lattices and fidelity ${\cal
F}(t)$ for the $N=12$ lattice.
\label{Fig_1}}
\end{figure}


\begin{figure} \centering
\resizebox{0.8\linewidth}{!} {\includegraphics{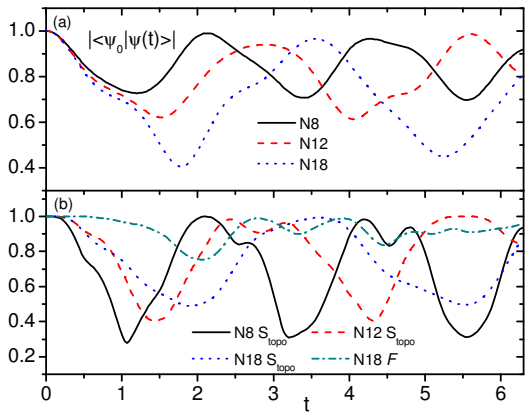}}
\caption{(Color online) Numerical results for the quench $H_4$
with $h = 0.34$. (a) Absolute overlap $|\langle \Psi_0 | \Psi(t)
\rangle |$ on lattices with $N=8,12,18$ spins. (b) $S_{\rm
topo}(t)$ for the $N=8,12,18$ lattices and fidelity ${\cal F}(t)$
for the $N=18$ lattice.
\label{Fig_2}}
\end{figure}

We are now ready to examine non-trivial quench Hamiltonians for
small lattices by means of exact numerical diagonalization. The
quantities of interest are $|\langle \Psi_0 | \Psi(t) \rangle |$,
$S_{\rm topo}(t)$ and ${\cal F}(t)$. We start with
\begin{eqnarray} \label{H_3}
H_{3}= - J_1\sum_{i} Z_i - J_2\sum_{\langle i,j \rangle} X_i X_j.
\end{eqnarray}
For $J_1  \gg J_2$ this Hamiltonian reduces to $H_1$, while for
$J_2  \gg J_1$ it reduces to $H_2$. We have shown that for both of
these extremes the TO of the initial state is robust during the
evolution. By contrast, in the intermediate regime, the recurrence
time increases with the size of the system. The two competing
terms affect both loop structures that support TO. The $S_{\rm
topo}$ and the ${\cal F}(t)$ quickly drop to low values (we note
that ${\cal F}(t)$ is typically more sensitive numerically than
$S_{\rm topo}$). The results are presented in Fig. \ref{Fig_1}.
The system undergoes an eigenstate thermalization in which TO is
destroyed. Having used several values of $J_1/J_2$ we find that
the most disrupting case for TO occurs for $J_1/J_2 \approx 0.33$,
which is the precursor of the quantum critical point for $H_3$. In
Ising-like models the speed of interactions is $\sim \sqrt{J_1
J_2}$ \cite{hamma_lrs} and is maximum at the critical point. At
this value correlations spread faster in the lattice and the
system samples a greater number of states in the Hilbert space.

Are there non-trivial quenches that preserve topological order? At
first sight this would seem unlikely because in such cases the
total state typically explores large parts of the Hilbert space.
However the Hamiltonian
\begin{eqnarray} \label{H_4}
H_{4} = -\sum_{s} A_{s} - \sum_{p} B_p - h \sum_{i} Z_i
\end{eqnarray}
displays interesting behavior. A reasonable scenario is one where
the system is prepared in the ground state of $H_4$ for $h=0$ and
then it is let to evolve for some $h \neq 0$. This can happen,
e.g., if the ground state of the TCM is precisely prepared using
syndrome measurements but there is some defect in the laboratory
implementation. The model was analyzed in Ref.
\cite{Hamma_detectionTQPT} and it was shown that there is a
precursor of a quantum phase transition even for small systems
(for $N=8$ we have $h_c \approx 0.34$). In particular, it was
shown there that for $h < h_{\rm c}$ the system is TO and for
$h>h_{\rm c}$ it is normally ordered.


\begin{figure} \centering
\resizebox{0.87\linewidth}{!} {\includegraphics{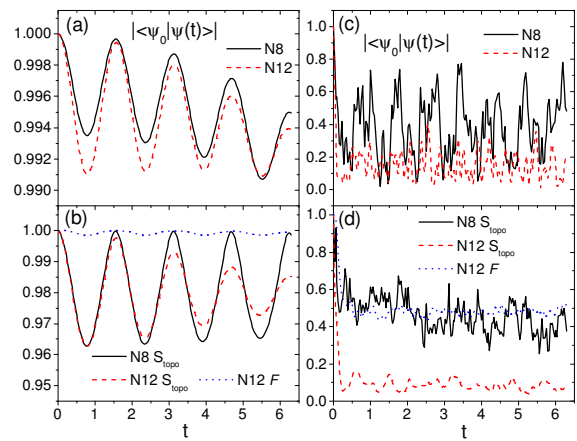}}
\caption{(Color online) Numerical results for the quench $H_5$.
(a) $|\langle \Psi_0 | \Psi(t) \rangle |$ on the $N=8,12$ lattices
for $J_1 = J_2 = 0.033$. (b) $S_{\rm topo}(t)$ for the $N=8,12$
lattices and ${\cal F}(t)$ for the $N=12$ lattice for $J_1 = J_2 =
0.033$. (c) The same as in (a) but for $J_1 = J_2 = 3.3$. (d) The
same as in (b) but for $J_1 = J_2 = 3.3$
\label{Fig_3}}
\end{figure}

We find that this quench is non-trivial: as the system size
increases the overlap $|\langle \Psi_0 | \Psi(t) \rangle |$
becomes smaller and recurrences occur later. In Fig. \ref{Fig_2}
we present the results for $h = 0.34$. The influence of this
quench on the system is milder than $H_3$ due to the fact that
$H_4$ couples the initial state to a much smaller Hilbert space as
it is constrained by $B_p=1$ for every plaquette $p$. On the other
hand, the time-averaged $S_{\rm topo}$ increases with the system
size. For small values of $h$ this is in line with the statical
behavior analyzed in Ref. \cite{Hamma_detectionTQPT}: the ground
state of $H_4$ for $h \ll 1$ has TO and $S_{\rm topo}$ approaches
the maximum value in the thermodynamic limit. Nevertheless, there
is an important difference with the statical case in that large
values of $h$ destroy TO (the ground state of $H_4$ for $h \gg 1$
is a trivial separable state). For the sudden quench, by contrast,
large values of $h$ preserve TO as the size of the lattice
increases.

We also present a non-trivial quench where the different possible
phases co-exist in the dynamical behavior,
\begin{eqnarray} \label{H_5}
H_{5} = -\sum_{s} A_{s} - \sum_{p} B_p-J_1\sum_{i} Z_i -
J_2\sum_{\langle i,j \rangle} X_i X_j.
\end{eqnarray}
For $J_1 \gg J_2 > 0.33$ or $J_2 \gg J_1 > 0.34$ the system
undergoes a periodic dynamic as with the quenches $H_1,H_2$. If
$J_2\ll J_1\sim 0.34$ and $J_1\ll J_2\sim 0.33$ the system has a
non-trivial dynamic behavior where TO is preserved as with $H_4$.
Finally we have the most interesting case when $J_1 \sim J_2$ (see
Fig. \ref{Fig_3}). Here we have two different phases: if $J_1 \gg
0.33$, the quench is destructive; but if $J_1 \ll 0.33$, TO is
preserved, which has been verified by checking that the
time-averaged topological entropy increases with lattice size.

In summary, we investigated how a topologically ordered ground
state reacts to rapid quenches. TO was detected by the topological
entropy and a new dynamical quantity, the fidelity between reduced
states belonging to different topological sectors. Clearly TO is
preserved if the quench Hamiltonian has a ground state belonging
to a topologically ordered phase. Nevertheless, we showed that
quenches in a different phase can still preserve TO. On the other
hand, we also showed that there is a class of quench Hamiltonian
(see Eq.(\ref{H_3})), whose dynamics leads to a suppression of TO.
The destructive dynamics results from Hamiltonians that are both
non-topologically ordered and have a finite propagation of
interactions through the lattice. It is remarkable that once a
state with TO has been initialized, it can be more robust with
respect to dynamical evolution, as opposed to statical
perturbations. It would be interesting to study the evolution of
the TCM in the presence of a suitable noise model under various
quenches. This would be a first step in constructing a dynamical
decoupling protocol \cite{dd} for the protection of the TCM from
thermal noise.

We thank P. Calabrese, J. Pachos, P. Zanardi for useful comments.
We acknowledge financial support by the EPSRC (D.I.T., grant no.
EP/D065305/1), the National Research Foundation and Ministry of
Education Singapore (R.F.), the NSF (S.H., grant no. DMR-0804914)
and the Perimeter Institute for Theoretical Physics. We also thank
the USC Center for High Performance Computing and Communications
for use of facilities.


\end{document}